\documentclass[graybox]{svmult}

\usepackage{type1cm}        
\usepackage{makeidx}         
\usepackage{graphicx}        
\usepackage{multicol}        
\usepackage[bottom]{footmisc}

\usepackage{cite}
\usepackage{newtxtext}       %
\usepackage[varvw]{newtxmath}       

\usepackage{enumerate}
\usepackage{amsmath}
\usepackage{epsfig}
\usepackage{latexsym}
\usepackage{xspace}

\usepackage{inputenc}
\usepackage{indentfirst}
\usepackage{color}
\usepackage{epstopdf}
\usepackage{colortbl}
\usepackage{blindtext}
\usepackage{lineno}
\usepackage{hyperref}
\usepackage{makeidx}

\makeindex             

\begin{document}


\title*{The quark-gluon plasma: diagnosis with thermal hadron production from the early history until detailed characterization at high energy colliders}  

\titlerunning{The quark-gluon plasma: diagnosis with thermal hadron production....}


\author{
Peter Braun-Munzinger, Krzysztof Redlich, and Johanna Stachel
}

\institute{Peter Braun-Munzinger \at
Extreme Matter Institute EMMI, GSI, 64291 Darmstadt, Germany and Physikalisches Institut, Universit\"{a}t Heidelberg, 69120 Heidelberg, Germany; 
\email{p.braun-munzinger@gsi.de}
\and
Krzysztof Redlich \at Institute of Theoretical Physics, University of Wroclaw, 50204 Wroclaw, Poland and Polish Academy of Sciences PAN, 50449 Wroclaw, Poland; 
\email{krzysztof.redlich@uwr.edu.pl}
\and
Johanna Stachel \at Physikalisches Institut, Universit\"{a}t Heidelberg, 69120 Heidelberg, Germany;\\ 
\email{stachel@physi.uni-heidelberg.de}
}

\maketitle


\abstract{
In nuclear collisions at relativistic energies, matter is created which resembles closely the matter that filled all space until about 15 $\mu s$ after the big bang. Here we summarize selected aspects of the research that led to the establishment of this new sub-field of physics and briefly describe its current `state of the art' with emphasis on matter creation through thermal particle production. In particular, we will focus on particle production at low transverse momentum and explain how its analysis sheds light on one of the key open questions, i.e. what is the mechanism of hadronization of colored objects such as quarks and gluons.
}

\section{Introduction and some historical context}
The story actually starts long before QCD, the quantum field theory of the strong interaction, was developed.  In the early 1950s it was realized that collisions between two protons in the GeV energy range led to multiple production of new particles (mostly pions). Their multiplicity exhibited features which could be understood in terms of a thermal approach ~\cite{Pomeranchuk:1951ey,Fermi:1953zz,Landau:1953gs}. With the advent of new higher energy accelerators at CERN (the CERN PS completed in 1959) and at Brookhaven (the BNL AGS completed in 1960) the stage was set to embark on a detailed study of how new particles are created in high energy proton-proton collisions. Very soon it was realized that some aspects of the experimentally emerging collision dynamics exhibited features characteristic of thermal particle production~\cite{Hagedorn:1965st,Hagedorn:1968jf,Hagedorn:1968zz}. Soon afterwards, other statistical approaches were published ~\cite{Bjorken:1969wi,Shuryak:1978ij}.

 In the early 1970s QCD was introduced with its novel features of infrared slavery and asymptotic freedom.  These properties of QCD  are outlined in detail in a recent review~\cite{Gross:2022hyw} and provide also the basis to describe  strongly interacting matter in terms of  thermodynamic phases. This was first realized by Collins~\cite{Collins:1974ky}. Soon thereafter appeared the 1st phase diagram of QCD ~\cite{Cabibbo:1975ig}. The term quark-gluon plasma (QGP) was actually coined soon afterwards by Shuryak~\cite{Shuryak:1977ut} for the description of the high temperature phase where confinement gives way to de-confinement and chiral symmetry is restored. Similar arguments apply also for matter at high baryon density such as is found in the center of neutron  stars.

This article is organized in the following way: we will first briefly summarize what is currently known about the QGP from first principle QCD. This is obtained by solving the QCD equations by introducing a discrete space-time lattice, i.e. using lattice QCD (lQCD). After a brief review of global quantities characterizing the QGP, the hadron resonance gas (HRG) is introduced. A good recent introduction into the physics behind the HRG is given in ~\cite{Ratti:2021ubw}. 

The key idea behind the HRG approach originates from the work of ~\cite{Hagedorn:1965st} and is the following: to derive thermal properties of a system of strongly interacting particles (hadrons or quarks and gluons) in equilibrium one needs the statistical partition function. In QCD this is directly obtained from the basic constituents (quarks and gluons) and their interaction encoded in the QCD Lagrangian. Since no analytic solution of the QCD equations is available to date a promising  alternative approach is to build the partition function from all known hadrons. Implicitly one can argue that, for temperatures not too far from the QCD phase transition temperature, the complete spectrum of hadrons already comprises all of QCD. Also, for such relatively low temperatures, a next step is then to neglect all interactions among hadrons. If one knows the masses and quantum numbers of all hadrons along with their strong decay probabilities it is then fairly straightforward to build the complete partition function. From this partition function one can compute all needed thermodynamic quantities such as energy density, entropy density and particle density. Explicit details are given in ~\cite{Braun-Munzinger:2003pwq,Andronic:2017pug,Ratti:2021ubw}.

An important refinement is to take (residual) interactions in the HRG description  via the S-matrix approach. This is briefly introduced in the section on S-matrix thermodynamics below. Another important development is the introduction of conserved quantum numbers  into the thermal approach. This leads directly to thermodynamical concepts for the QGP and forms the basis of the statistical hadronization model (SHM) when coupled to a prescription on how hadrons are formed at the QCD phase boundary. This prescription essentially follows the ideas developed in ~\cite{Cooper:1974mv}. 

We will dominantly focus on the global properties of integrated yields of particles produced in hadron and heavy ion collisions. This will allow us to address the concept of their thermal origin and to identify the equation of state of the fireball produced in the collision. 
This is particularly transparent at the highest collision energies at the LHC, where measured yields of light-quark hadrons are shown to be consistent with the lQCD equation of state at the QCD phase boundary.     

Identification of the thermal origin of hadron production is therefore the prerequisite before proceeding to the study of interesting collective and dynamical phenomena outside the scope of the SHM. Also the majority of studies of fluctuations and correlations as well as investigations of  critical properties linked to the chiral phase transition occuring in hot and dense QCD matter \cite{Chen:2024aom} are performed within the framework of statistical theories~\cite{Stephanov:1999zu,Braun-Munzinger:2015hba}. 

A high degree of thermal equilibrium is equally necessary to analyze the expansion dynamics of the fireball \cite{Gale:2013da}, and its transverse and elliptic flow \cite{Snellings:2011sz,Ollitrault:2009ie} with the tools of relativistic hydrodynamics.  Investigations into the microscopic origin of parton energy-loss \cite{Baier:1996sk} and  experimental and theoretical studies of electromagnetic emissivity in heavy-ion collisions have brought  interesting insights into  properties of the QGP medium~\cite{Rapp:2013nxa,Rapp:2009yu}. 

Many of these interesting aspects are outside of the scope of the present article and are therefore only mentioned and cited here. We do discuss below in some detail the case of multi-charm hadrons and their special role in the QGP as it is presently our only means to shed light on the crucial question of the degree of deconfinement of the medium under investigation ~\cite{Andronic:2017pug,Gross:2022hyw}.


We  demonstrate the predictive power of the SHM for the description of experimental particle yields for hadrons composed of light (u,d,s) valence quarks. For an extension of SHM to heavier quarks we introduce the concept of the heavy quark (charm or beauty) balance equation and demonstrate that this leads to a parameter-free extension to the charm sector (SHMc) as demonstrated by comparing to relevant experimental data. We follow with a short section on the possible role of charm quarks in the QGP equation of state and end with a brief summary. 

\section{Short summary of recent results from lQCD}

In the thermodynamic limit, i.e under the assumption of thermal equilibrium, the full QCD partition function, and hence the equation of state, can be computed without further assumptions, although limits from computer technology initially did not allow inclusion of dynamical quarks.
The first lQCD calculations of the equation of state were performed in the early 1980s ~\cite{Engels:1980ty}. Already then a close link emerged between deconfinement and restoration of chiral symmetry ~\cite{Kogut:1982rt}. With increasing computing power the following four decades brought a real breakthrough in finite temperature lQCD. 

Recent numerical lQCD calculations~\cite{HotQCD:2019xnw} show, for massless u and d quarks and at vanishing chemical potential, evidence for a true second-order chiral transition between hadron gas and QGP at a critical temperature of $T_{c}\approx132^{+3}_{-6}$ MeV. For realistic u,d,s-quark masses, the chiral transition is rather a crossover at a precisely determined  pseudo-critical temperature of $T_{pc}$ = 156.5 $\pm 1.5$ MeV ~\cite{HotQCD:2018pds}. Independent confirmation came from ~\cite{Borsanyi:2020fev}, where a transition temperature of 158.0 $\pm$ 0.6 MeV was reported. For a very recent review, see ~\cite{Aarts:2023vsf}.

Extending the lQCD calculations into regions of finite net baryon density quantified by a baryon chemical potential $\mu_B$~\cite{HotQCD:2018pds,Borsanyi:2020fev} is now possible by expansion techniques. Current lQCD expansion techniques already cover the regime of $\mu_B/T\leq 3$. The so obtained phase transition line of pseudo-critical temperatures is shown in the QCD phase diagram displayed in Fig~\ref{fig:phase_diagram} below. The sign problem so far prevents use of this technique in regions of larger values of $\mu_B$, see e.g. ~\cite{Gattringer:2016kco}. There one has to resort to models of QCD for further exploration. All indications point since a long time to a very similar temperatures and phase transition line for the deconfinement transition.

\section{Brief summary of experimental results  on global QGP properties}

Experimentally, this regime of the QCD phase transitions is accessible by investigating collisions of heavy nuclei at high energy. It was conjectured already in~\cite{Shuryak:1978ij} that, in such hadronic collisions, after some time local thermal equilibrium is established and all properties of the system (fireball) are determined by a single parameter, the temperature $T$, depending on time and spatial coordinates. This is exactly the regime probed by collisions of nuclei at the Large Hadron Collider (LHC), as will be outlined in the following. The region of finite to large $\mu_B$ is accessed by nuclear collisions at lower energies.

In the early phase of the collision, the incoming nuclei lose a large fraction of their energy leading to the creation of a hot fireball characterized by an energy density $\epsilon$ and a temperature $T$. The concomitant deceleration of the nucleons in the colliding nuclei, called stopping, is characterized by an average rapidity shift $\Delta y$ = -ln($E/E_0$) with  nucleon energies $E$ and $E_0$ before and after the collision. Quantitative information is contained in the experimentally measured net-proton rapidity distributions (i.e. the difference between proton and anti-proton rapidity distributions). These distributions are summarized for collision energies from the SPS to RHIC energy range in~\cite{Braun-Munzinger:2020jbk}. Including also the AGS data it emerges that the rapidity shift saturates at approximately two units from $\sqrt{s_{NN}}\approx$ 17.3 GeV upwards, implying a fractional energy loss of $1-{\rm exp}(-\Delta y) \approx 86\%$. In fact, the same rapidity shift was already determined for p--nucleus collisions at Fermilab for 200 GeV/c proton momentum~\cite{Abe:1988hq} compared to about one unit for pp collisions. With increasing collision energy, the target and projectile rapidity ranges are well separated, leaving at central rapidity with a small or even zero 
net-baryon density, and universal fragmentation regions forward and backward following the concept of limiting fragmentation~\cite{Benecke:1969sh}.

The energy loss (rapidity shift) of the incident nucleons leads to high energy densities at central rapidity, i.e., in the center of the fireball. These initial energy densities can be estimated using the Bjorken model~\cite{Bjorken:1982qr}:
\begin{equation}
\label{eq-Bj}
    \epsilon_{BJ} = \frac{1}{A\tau_{0}}\frac{dE_{T}}{d\eta}\frac{d\eta}{dy},
\end{equation}
where $A=\pi r^{2}$ is the overlap area of two nuclei and the kinetic equilibration time $\tau_0$. Eq.~\ref{eq-Bj} is typically evaluated at a time $\tau_{0}$ = 1 fm and the resulting energy densities are displayed in Table~\ref{tab:Edensity} for central Au--Au and Pb--Pb collisions at the different collision energies. For central Pb--Pb collisions ($A =$ 150 fm$^2$) at $\sqrt{s_{NN}}$ = 2.76 TeV this yields an energy density of about 14 GeV/${\rm fm^3}$~\cite{CMS:2012krf}, more than a factor of 30 above the critical energy density for the chiral phase transition as determined in lQCD calculations. In fact, for all collision energies shown the initial energy density significantly exceeds that computed in lQCD at the pseudo-critical temperature, indicating that the matter in the fireball is to be described with colored quark and gluon degrees of freedom rather than as hadronic matter. The corresponding initial tem\-pe\-ra\-tures can be computed using the energy density of a gas of quarks and gluons with two quark flavors, $\epsilon = 37\frac{\pi^{2}}{30}T^{4}$, yielding e.g. $T\approx$ 307 MeV for LHC energy. Temperature values for lower collision energies are also quoted in the Table\footnote{The values reported in the table are all for vanishing chemical potentials. We have evaluated the differences if one assumes values for chemical potentials as determined at chemical freeze-out, see below. The resulting temperature values differ by less than 5\%  from those reported in Table~\ref{tab:Edensity}. Owing to the proportionality of energy density to the fourth power of temperature, inclusion of a bag pressure only mildly changes the calculated temperature values.}.

\begin{table}[htb]
\centering
\begin{tabular}{| c | c | c | c | c |}
 \hline
 & $\sqrt{s_{NN}}$  & $dE_{t}/d\eta$ & $\epsilon_{BJ}$ & T \\ [2ex]
 & [GeV] & [GeV] & [GeV/$fm^{3}$] & [GeV] \\ 
 \hline
AGS & 4.8 & 200 & 1.9 & 0.180 \\  [1ex]
 \hline
SPS & 17.2 & 400 & 3.5 & 0.212 \\ [1ex]
\hline
RHIC & 200 & 600 & 5.5 & 0.239 \\ [1ex]
\hline
LHC & 2760 & 2000 & 14.5 & 0.307 \\ [1ex]
\hline
\end{tabular}
\caption{Collision energy per colliding nucleon pair, measured transverse energy pseudo-rapidity density at mid-rapidity~\cite{E814E877:1993rlr,WA98:2000mvt,PHENIX:2015tbb,CMS:2012krf}, energy density, and initial temperature estimated as described in the text for central Pb--Pb and Au--Au collisions at different accelerators.}
\label{tab:Edensity}
\end{table}

Depending on energy, collisions of heavy ions populate different regimes falling into two categories: (i) the stopping or high baryon density region reached at $\sqrt{s_{NN}} \approx$ 3-20 GeV and (ii) the transparency or baryon-free region reached at higher collision energies. The net-baryon-free QGP presumably existed in the early universe after the electro-weak phase transition and up to about 15 microseconds after the Big Bang. This corresponds to the time of the QCD phase transition, where the strongly interacting constituents, i.e. the quarks and gluons, are converted into hadrons. For a brief but concise description, see section 22.3 of ~\cite{ParticleDataGroup:2024cfk}. In the QGP of the early universe, particles interacting via the strong and electro-weak force are part of the system, while an accelerator-made QGP only contains strongly interacting particles. On the other hand, a baryon-rich QGP may be produced in neutron star mergers or could exist, at very low temperatures, in the center of neutron stars~\cite{Bauswein:2018bma,Baym:2019iky,Gorda:2022jvk}. 

Another important difference between the `laboratory-created' QGP and the QGP phase in the early universe is that, after the QCD phase transition, the `laboratory-created' QGP falls out of equilibrium and never returns. This is denoted as chemical freeze-out. In contradistinction, the standard model phase, which now contains hadrons, leptons, photons and neutrinos, immediately returns to equilibrium and stays there until neutrino-freeze-out at a time of about 1 s after the big bang. For a detailed description of this phase see ~\cite{Rafelski:2023emw}. 

\section{S-matrix thermodynamics of strong interactions}

A fireball produced in particle or nuclear collisions can be described as a thermal medium composed of hadrons as leading constituents.  The thermodynamics of such a system is described in terms of an interacting gas of ground-state hadrons. The S-matrix formalism is a theoretical framework to implement interactions in a dilute many-body hadronic system in thermal and chemical equilibrium \cite{Dashen:1969ep,Venugopalan:1992hy,Weinhold:1997ig,Lo:2017lym,Lo:2020phg}.  There, the thermodynamic pressure for a hadronic gas at finite temperature and density, formulated in the Grand-Canonical (GC) ensemble, undergoing two-body scatterings, $h_1+h_2\to h_1+h_2$, is expressed as
\begin{equation}
\label{eqn:pressure0}
P(T,\vec \mu)=\sum_iP_{i}^{id}+\Delta P^{int},
\end{equation}
where $P_{i}^{id}$ corresponds to an ideal gas contribution of hadron $h_i$.
The interaction  part of  the thermodynamic pressure due to  two-body scatterings $\Delta P^{int}$
involves an integral over the invariant mass $M$~\cite{Weinhold:1997ig},

\begin{equation}
\label{eqn:pressure}
    \Delta P_{int.} 
   \approx \sum_{I,j}  \int_{m_{th}}^\infty  dM \frac{1}{\pi}  \frac{d \delta^I_j}{dM}  P^{id}(M)  \\
\end{equation}
\noindent where $\delta^I_j$ is the scattering phase shift for a given isospin-spin channel.

The pressure of a non-interacting gas of particles and their anti-particles carrying mass $m_i$ and spin-isospin degeneracy factor $d_i$, under Boltzmann statistics, reads\footnote{We use the Boltzmann distribution for simplicity. The extension to quantum statistics can be found in Refs. \cite{Braun-Munzinger:2003pwq,Lo:2017lym} and is applied when doing quantitative investigations.}   
\begin{equation}
	\label{eqn:idp}
	P^{id}_{i}= \frac{d_i}{2\pi^2} m_i^2 T^2 \left( \lambda_i + \frac{1}{\lambda_i} \right) K_2\left( \frac{m_i}{T} \right),    
\end{equation}
where $\lambda_i=\exp[(B_i\mu_B+S_i\mu_S+Q_i\mu_Q)/T]$ is the fugacity,  and $\vec\mu =(\mu_B,\mu_S,\mu_Q)$ is the chemical potential linked to the GC implementation of charge-conservation laws with $B_i,S_i,Q_i$ being the baryon number, strangeness and electric charge of particle, $i$, respectively.     

In the presence of attractive interactions driven by a resonance of mass $m$ and vanishing width, $\Gamma\simeq 0$, the effective density of states 
approaches a delta function,
\begin{equation}
\label{eqn:tr}
\frac{1}{\pi}  \frac{d \delta^I_j}{dM}\to \delta(M-m).
\end{equation}
Consequently, the 
interaction part of the pressure acts as an ideal gas of resonances.  
For a  finite resonance width,  the spectral density  is linked to  the standard Breit-Wigner distribution, 
\begin{equation}
	\label{eqn:smat2bw}
	\frac{d}{d M} \delta_{\rm res}(M) 
		       \to B(M)=\frac{2 M^2 \Gamma}{(M^2-{m_{}}^2)^2 + M^2 \Gamma^2}.
\end{equation}
In the limit $\Gamma\to 0$ the spectral density converges to Eq.~\ref{eqn:tr}.

Thus, in leading order in the density expansion, accounting only for attractive interactions driven by resonance formation, the thermodynamic pressure of strongly interacting hadron gas can be  approximated as a mixture of ideal gases of all stable hadrons, resonances, and their antiparticles, 
\begin{equation}
	\label{HRG}
	P^{HRG}(T,\vec \mu)=\sum_i P^{id}_{i} +\sum_k P^{res}_k.      
\end{equation}
The first sum runs over all stable hadrons, whereas the second term quantifies the contribution of all known resonance states, where
\begin{equation}
\label{pres}
    P^{res}_k= 
     \int_{m_{th}}^\infty {{dM}\over \pi} B(M)P^{id}_k  . \\
\end{equation}
This approximation of the S-matrix constitutes the well-known Hadron Resonance Gas (HRG) model \cite{Braun-Munzinger:2003pwq,Andronic:2017pug}. 

\subsection{Hagedorn's limiting  temperature}
The HRG thermodynamic potential in Eq.\ref{HRG} is a discrete representation of the statistical-thermodynamical approach to strong interactions at high energies formulated by Rolf Hagedorn in 1965, based on the  
concept of the statistical bootstrap \cite{Hagedorn:1965st, Hagedorn:1971mc} (see also \cite{Rafelski:2016nxx} and references therein). He assumed that, at high collision energies, higher and higher resonances (fireballs) of strongly interacting hadrons are formed and take part in the thermodynamics as if they were stable particles. A self-similar scheme for the composition and decay of resonances led Hagedorn to the solution of their  exponential mass spectrum, 
\begin{equation}
\label{TH}
\rho(m)=const\cdot m^{a}\exp(m/T_H),
\end{equation}
where $a=-5/2$ \cite{Hagedorn:1965st}. Consequently, the pressure of resonances corresponding to the above $\rho(m)$ becomes  \cite{Hagedorn:1971mc}     
\begin{equation}
\label{hag}
    P_H(T)
    \simeq {T^2\over{2\pi^2}} 
     \int_{m_{0}}^\infty dm ~m^{2}\rho(m) K_2({m\over T})  \\
\end{equation}
The exponential form of $\rho(m)$ implies that $T_H$ becomes the upper limit of permissible temperature of hadronic matter. See also the discussion about this in ~\cite{Huang:1970iq}. The power law coefficient $a$ determines the behavior of different thermodynamic quantities at $T\simeq T_H$. With the value of $a=-5/2$ obtained by Hagedorn, the pressure, energy density and the specific heat diverge at $T_H$. This is what made Hagedorn conclude that $T_H$, known today as Hagedorn's temperature, is indeed the highest possible temperature of hadronic matter. On the other hand, for a lower value of $a$, in particular, for $a=-7/2$ as expected e.g. from the Veneziano model \cite{Fubini:1969qb}, the energy density remains finite at $T_H$, shifting the divergence to the specific heat \cite{Satz:1978us}.
Also, formulating Hagedorn's model for extended particles can provide a finite energy density at $T_H$ for larger values of $a$ \cite{Hagedorn:1980kb}. Such a behavior of thermodynamic quantities is usually linked to the transition to 
a new phase of matter. Today, one  connects this to the transition from hadronic matter to a quark-gluon plasma. Furthermore, the value of the Hagedorn temperature, extracted from different implementations of the bootstrap conditions \cite{Hagedorn:1971mc,Rafelski:2016nxx}, or from the recent fit of  $\rho(m)$ to data \cite{Lo:2015cca,ManLo:2016pgd,Broniowski:2000bj}, lies in the range $135<T_H<190$ MeV. This range contains the precisely determined chiral crossover temperature $T_c\simeq 156.5$ MeV,  calculated within the framework of lQCD (see above).       

\subsection{Linking the HRG with the QCD thermodynamics of the hadronic phase}
A long-lasting question was whether the HRG model approximation of the thermodynamic potential of the hadronic phase that accounts for only attractive interactions is consistent with QCD thermodynamics. The answer and justification were given by directly comparing the equation of state obtained in lQCD and from the HRG~ \cite{Karsch:2003vd,Karsch:2003zq,Ejiri:2005wq,Allton:2005gk,HotQCD:2012fhj,Karsch:2013naa,Bazavov:2017dus,Bellwied:2021nrt}, see also \cite{Karsch:2022jwp} or Hagedorn's model \cite{Lo:2015cca,Alba:2017mqu}. It turned out that the temperature dependence of pressure, energy density and entropy density is well described by the HRG equation of state up to the near vicinity of the chiral crossover. Such an agreement is direct evidence that the thermodynamic potential of the HRG already provides a good approximation to QCD thermodynamics of the confined phase. Interestingly, by comparing HRG model results  with lQCD results, it was shown that there must be missing resonances beyond those listed in the PDG~\cite{ParticleDataGroup:2024cfk},
particularly in the strange-baryon and charmed-baryon sectors \cite{Alba:2017mqu,Bollweg:2022fqq,Sharma:2025zhe}.  

\subsection{Statistical Hadronization Model and hadron production yields}
From the partition function of HRG  
it is rather straightforward to calculate the particle composition of a fireball of volume $V$ and with thermal 
parameters $(T, \vec{\mu})$ \cite{Braun-Munzinger:2003pwq,Andronic:2017pug,Braun-Munzinger:1999hun}. The  total multiplicity $\langle N_i\rangle$ of particle  $i$ is obtained, as
\begin{equation}
\label{yieldR}
\langle N_i\rangle
= \langle N_i\rangle^T + \sum_r B(r\rightarrow i) \langle N_r\rangle^{res}. 
\end{equation}
The first term is the thermal average number of particles $i$. The second term describes the overall contribution from resonances decaying to $i$. The $B(r\to i)$ is the corresponding decay branching of resonance $r$ to final state hadron $i$. In the Boltzmann approximation, enter the thermal yields of stable particles, 
$\langle N_i\rangle^T=(V/T)P^{id}_i$ with $P^{id}_i$ from Eq. \ref{eqn:pressure0}, and of resonances $\langle N_r\rangle^{res}=(V/T)P^{res}_r$ with $P^{res}_r$ as in   Eq. \ref{pres}. 

The numerical implementation of Eqs.~(\ref{HRG}-\ref{yieldR}), extended to quantum statistics,   
constitute the Statistical Hadronization Model (SHM), as given in ~\cite{Andronic:2017pug} and refs. there. This model has been successfully applied to quantify thermalization and particle production yields in hadron and heavy ion collisions (see the discussion below).  

We note, however, that 
some of the simplifying assumptions as e.g. that of attractive interactions implicit in the above HRG model are not necessarily 
consistent with hadron scattering data. In particular, for an accurate implementation of interaction effects, a proper resonance invariant mass distribution, the presence of 
non-resonant contributions, as well as repulsive interactions, have to be included to be consistent with scattering data.
This can be done systematically within the S-matrix approach where two-body interactions are included via the empirical scattering phase shifts as in Eq. \ref{eqn:pressure}. The resulting interacting density of states is then folded into an integral over thermodynamic distribution functions which, in turn, yields the contribution from interactions to a particular thermodynamic quantity.

The phenomenological implications and importance of the S-matrix approach have been recently quantified in the context of particle production in heavy ion collisions, and in the description of lQCD results on different fluctuation and correlation observables of conserved charges in the hadronic phase. 
In particular, it was demonstrated that the implementation of the empirical pion-nucleon phase shifts is crucial for the quantitative description of lqcd results on the baryon-charge susceptibility  \cite{Lo:2017lym}.  Also in the analysis of proton production yields in nucleus-nucleus collisions at the LHC, a careful treatment of empirical pion-nucleon scattering phase-shifts could resolve the proton-yield anomaly \cite{Andronic:2018qqt}. Furthermore, in the strange baryon sector of the HRG, the improvement of interactions within the S-matrix formalism was shown to increase the strange-baryon correlations towards lQCD results up to $T\simeq T_{c}$ \cite{Fernandez-Ramirez:2018vzu,Friman:2015zua}.

\subsection{Exact charge conservation}
The  HRG model for particle production introduced in  Eq.~\ref{HRG} is formulated in the  Grand Canonical (GC) ensemble concerning charge conservation laws. The conservation of baryon number, strangeness, and electric charge holds on average and is controlled by the corresponding chemical potentials. It is already well established, however, that such a GC model can only be applied if the number of produced charged particles is sufficiently large. If this is not the case, a thermal description requires exact implementation of charge conservation, as introduced in the canonical C-ensemble~\cite{Hagedorn:1971mc,Redlich:1979bf,Hagedorn:1984uy,Cleymans:1990mn,Cleymans:1998yb,Hamieh:2000tk,Ko:2000vp,Braun-Munzinger:2003pwq}. This is particularly the case when applying the thermal model to particle production in pp and pA collisions, as well as in low-energy heavy-ion collisions and even at the  LHC when considering events with low charged-particle multiplicity. In general, in collisions the conservation of quantum numbers is fulfilled exactly as the initial conditions fix them.    

The yields of charged particles calculated in the C-ensemble are usually suppressed relative to the values obtained in the GC-ensemble
\cite{Hagedorn:1984uy,Cleymans:1998yb,Braun-Munzinger:2003pwq,Rafelski:1984mq}.
The HRG model formulated in the C-ensemble has provided an instrumental framework for the centrality and system-size dependence of particle production and in particular strangeness production and suppression  \cite{Cleymans:1998yb,Hamieh:2000tk,Redlich:2001kb}. The
applicability of the model in small systems like pp \cite{Kraus:2008fh,Becattini:1997rv}, pA \cite{Sharma:2018owb} and e$^+$e$^-$ collisions has been successfully verified in the literature \cite{Andronic:2008ev,Becattini:2008tx}.

In particular, the characteristic prediction of the HRG model in the C-ensemble was an increasing suppression of strange particle yields per pion with decreasing collision energy and collision centrality, as well
as with increasing strangeness content of the particle \cite{Hamieh:2000tk,Redlich:2001kb}.
Such a pattern of suppression of (multi-)strange hadrons with decreasing multiplicity was indeed observed by the ALICE collaboration ~\cite{ALICE:2016fzo} and is qualitatively similar to what has been measured previously by the WA97 and NA57  collaborations at SPS energies at CERN \cite{WA97:2001xtk} and by the STAR collaboration at RHIC~\cite{STAR:2010yyv}.
The recent ALICE data on (multi)strange particle production, as well as data on kaon excitation functions at lower energies, are consistent with predictions of the thermal model  accounting for exact strangeness conservation formulated in the canonical ensemble \cite{Cleymans:2020fsc,Vislavicius:2016rwi,Cleymans:2000ck}.

\section{Hadron production at the QCD phase boundary: a contemporary view}

\subsection{Integrated yields (u,d,s) hadrons}
Experimental information on the equation of state of QCD matter near the phase boundary, and hence on the QCD phase diagram, is obtained from the measurement of the yields of hadrons produced in (central) high energy nuclear collisions. Analysis of  data in the framework of the SHM  makes use of Eq. \ref{yieldR},
see ~\cite{Andronic:2017pug} and refs. there,
since there it is demonstrated that at hadronization, when the colored quarks and gluons are transformed into colorless hadrons, the fireball formed in the collision is very close to a thermodynamic state in full (hadro-)chemical equilibrium. 

From Eq. \ref{yieldR}
one obtains directly the first moments (mean values) of the densities of hadrons as a function of a pair of thermodynamic parameters, the temperature $T_\mathrm{chem}$ and the baryon chemical potential $\mu_B$ at chemical freeze-out. Note that after chemical freeze-out, the fireball is rapidly falling out of equilibrium.  Expansion and cooling will move it further away from equilibrium~\cite{Braun-Munzinger:2003htr}. This situation is very different from the evolution in the early universe where the cooling rate is very slow compared to typical hadronic interaction times of order $10^{-23}$s and the system continuously adjusts to thermal equilibrium for an extended period.

\begin{figure}[!htb]
    \centering
    \includegraphics[width=0.65\linewidth,clip=true]{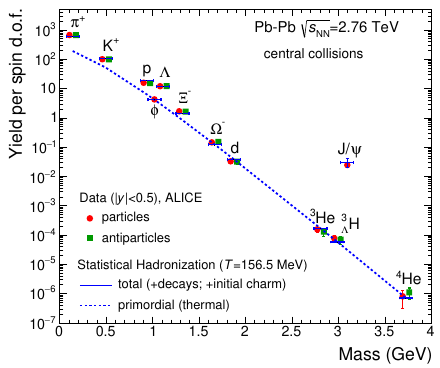}
    \caption{Experimental rapidity densities normalized to the spin degeneracy for LHC data compared to primordial and total (anti-)particle yields as calculated within the SHM~\cite{Andronic:2017pug}.}
    \label{fig:HRG}
\end{figure}

To go beyond the non-interacting HRG gas approximation, attractive and repulsive interactions between hadrons can be taken into account in the S-matrix formulation of statistical mechanics, as discussed above. The predictions of the SHM for hadron yields, including the S-matrix correction for protons, are compared to experimental data at LHC energy for $T_\mathrm{chem}  = 156.5$ MeV in Fig.~\ref{fig:HRG}. The agreement between model predictions and data is excellent for all measured hadron species, including nuclei and hyper-nuclei. Since the value of the baryo-chemical potential turns out to be consistent with zero, all anti-particle yields agree with those of the corresponding particles, with yields varying over 9 orders of magnitude. Remarkably, the description works also for loosely bound states, i.e. atomic nuclei where the nucleon separation energy is much smaller than the temperature of the system. 

\begin{figure}[!htb]
    \centering
    \includegraphics[width=0.6\linewidth,clip=true]{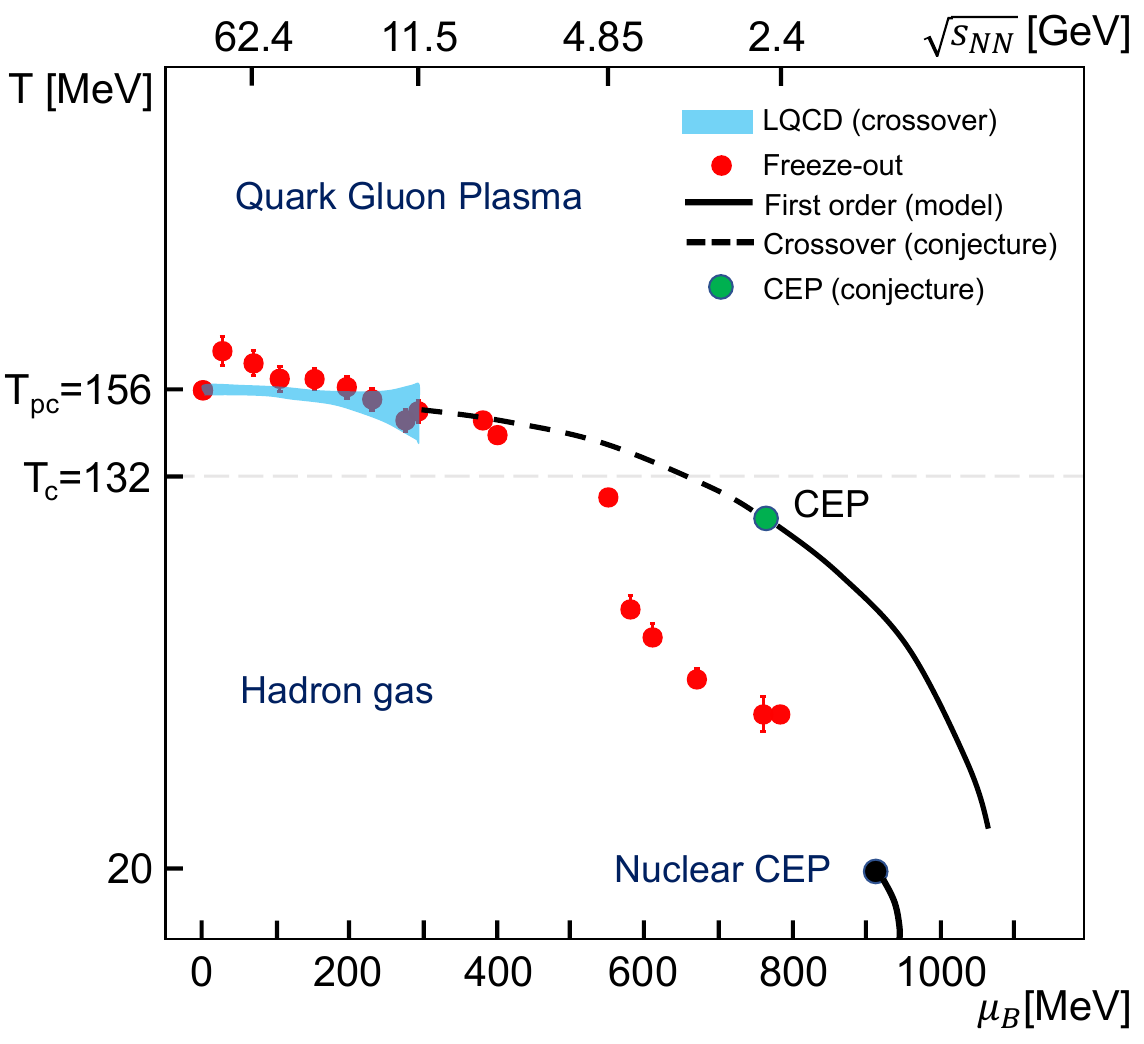}
    \caption{Phase diagram of strongly interacting matter. The red symbols correspond to chemical-freezeout parameters, temperature  $T_\mathrm{chem}$ and baryon chemical potential $\mu_{B}$ determined from experimental hadron yields~\cite{Andronic:2017pug, STAR:2017sal, HADES_FO}. The blue band represents the results of lQCD computations of the chiral phase boundary~\cite{HotQCD:2018pds,Borsanyi:2020fev}. Also shown are a conjectured line of first order phase transition with a critical end point as well as the nuclear liquid-gas phase boundary. This figure is from ~\cite{Gross:2022hyw}.}.
    \label{fig:phase_diagram}
\end{figure}

The values of the hadro-chemical freeze-out parameters determined through measurements at lower collisions energies are likewise obtained by fitting the SHM predictions to the measured hadron yields. The so extracted freeze-out parameters $T_\mathrm{chem}$ and $\mu_{B}$~\cite{Andronic:2017pug, STAR:2017sal, HADES_FO} are presented as red symbols in the QCD phase diagram shown in Fig.~\ref{fig:phase_diagram}. The experimental freeze-out points are compared to the crossover chiral phase transition line as computed in lQCD (shown as the blue band). For all center-of mass energies from LHC energies on down to about $\sqrt{s_{\mathrm{NN}}}$ = 12 GeV, i.e., over the entire range covered by lQCD, there is a remarkably close agreement visible between $T_\mathrm{chem}$ and the pseudo-critical temperature for the chiral cross over transition $T_{pc}$. We note that, along this phase boundary, the energy density computed (for 2 quark flavors) from the values of $T_\mathrm{chem}$ and $\mu_B$ exhibits a nearly constant value of $\epsilon_\mathrm{crit} \approx 0.46$ {GeV/fm}$^3$. 

The experimental result that the chemical freeze-out temperature turns out to be close to $T_{pc}$ has a fundamental reason: because of the very rapid temperature and density change across the phase transition and afterwards due to the rapid expansion, the resulting low hadron densities in the fireball combined with its size cause inelastic interaction between hadrons to cease within a narrow temperature interval of a few MeV~\cite{Braun-Munzinger:2003htr} after hadron formation. 

As already indicated above, this rapid chemical freeze-out is actually very different from  the continuous slow decoupling taking place in the early universe.  To emphasize the different time and distance scales relevant there, we note that for temperatures $T$ larger than a few MeV even the mean free path for neutrinos is much smaller than the size of the universe, see section 22.3 of ~\cite{ParticleDataGroup:2024cfk}. The slow decoupling after the QCD phase transition implies baryon-antibaryon annihilation and, due to the very small initial baryon-antibaryon asymmetry, to an increasing baryon chemical potential. The trajectory of the early universe in the QCD phase diagram thereby 
takes a somewhat unexpected course ~\cite{Braun-Munzinger:2008szb}. See also the discussion at the end of section 3 above. Due to this baryon asymmetry (still not understood, for a discussion see section 22.3.6 in ~\cite{ParticleDataGroup:2018ovx}), annihilation in the early universe stops when all antibaryons are `consumed'. The surviving protons and neutrons form nuclei, in competition with neutron decay, when the temperature has reached the MeV scale. The final outcome of this is described well in Steven Weinberg's textbook `The First Three Minutes' ~\cite{Weinberg:1977ji}.

For values of baryon chemical potential beyond 400 MeV, implying higher baryon density beyond the current reach of lQCD, experimental data for chemical freeze-out have also been measured but the resulting phase structure of strongly interacting matter at high baryon density is still not well understood. Calculations based on various model approaches imply the appearance of a line of first order phase transition. Combined with the crossover transition at smaller values of the baryon chemical potential this would imply the existence of a critical end point (CEP) in the QCD phase diagram as indicated in Fig.~\ref{fig:phase_diagram}. The experimental discovery of the CEP would mark a major break-through in our understanding of the QCD phase structure. If it exists, the location of the CEP is most likely in the region $\mu_B > 500$ MeV based on functional methods applied to QCD~\cite{Fu:2019hdw} and extrapolations from lQCD. Searching for the CEP is the subject of a very active research program, at RHIC and the future FAIR facility at GSI ~\cite{Bzdak:2019pkr}. 

\subsection{Extension to the heavy quark sector}
With the impressive progress in particle identification of the four large LHC experiments  significant new  information is obtained on particle production in  relativistic nuclear collisions. In particular, there is much progress for hadrons with open and hidden charm and beauty. In addition, there is mounting evidence ~\cite{ALICE:2022wpn,Andronic:2021erx,Andronic:2019wva}, that hadrons composed of charm quarks reach a large degree of thermal equilibrium, although charm quarks in the system initially are chemically far out of equilibrium. This is supported by heavy quark diffusion coefficients from lQCD~\cite{Altenkort:2020fgs}. A strong indication for equilibration is the fact that J/$\psi$ mesons participate in the collective, anisotropic hydrodynamic expansion ~\cite{ALICE:2013xna,He:2021zej}.

To microscopically understand the production me\-cha\-nism of charmed hadrons for systems ranging from pp to Pb--Pb, various forms of quark coalescence mo\-dels have been developed ~\cite{Cho:2019lxb,Zhao:2018jlw,ExHIC:2017smd,Zhou:2014kka,Greco:2003vf}. This provides one way to study the dependence of production yields on hadron size and, hence, may help to settle the still open question whether the many exotic hadrons that have been observed recently are compact multi-quark states or hadronic molecules (see ~\cite{Aarts:2016hap,Maiani:2022psl} and refs. cited there). Conceptual difficulties with this approach are that energy is not conserved in the coalescence process and that color neutralization at hadronization requires additional assumptions about quark correlations in the QGP~\cite{Song:2021mvc}.

\begin{figure}[!htb]
    \centering
    \includegraphics[width=0.65\linewidth,clip=true]{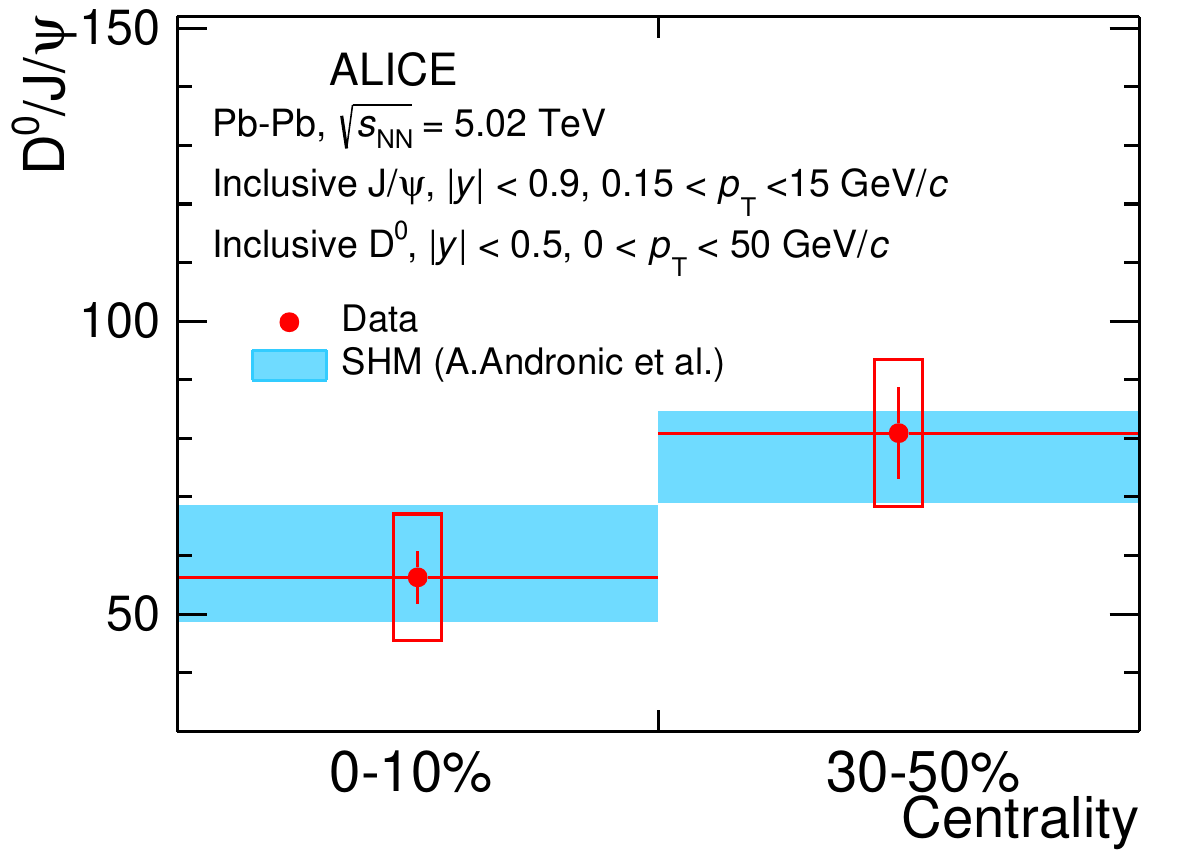}
    \caption{$D^{0}$ to J/$\psi$ yield ratio measured in Pb--Pb collisions at the LHC and predicted by the Statistical Hadronization Model with charm SHMc. Figure from~\cite{ALICE:2022new}.}
    \label{fig:DotoJpsi}
\end{figure}

Our own approach, named SHMc, has been made possible by the extension of the SHM to also incorporate charm quarks. This essentially parameter-free was first proposed in ~\cite{Braun-Munzinger:2000csl} and developed further in ~\cite{Andronic:2002pj,Grandchamp:2003uw,Becattini:2005hb,Andronic:2006ky,Andronic:2017pug,Braun-Munzinger:2024ybd}. A comprehensive discussion is given in ~\cite{Andronic:2021erx} and includes in SHMc all presently known hadrons with hidden and open charm. Importantly, the role of charmonia is not special anymore: they are treated as all other charmed hadrons. The key idea behind this is the recognition that, contrary to what happens in the (u,d,s) sector, the heavy (mass $\sim$ 1.2 GeV) charm quarks are not thermally produced. Rather, production takes place in initial hard collisions. The produced charm quarks then thermalize in the hot fireball, but the total number of charm quarks is conserved during the evolution of the fireball~\cite{Andronic:2006ky} since charm quark annihilation is very small. 

In essence, in this special thermal approach charm quarks can be treated like impurities in the hot fireball. Their description then requires the introduction of a charm fugacity $g_c$~\cite{Braun-Munzinger:2000csl,Andronic:2021erx}. The value of $g_c$ is not a free parameter but experimentally determined by measurement of the total charm cross section. For central Pb--Pb collisions at LHC energy, $g_c \approx 30$~\cite{Andronic:2021erx}. An important consequence of the large fugacity is a  strong increase in charmed hadron production compared to predictions from a purely thermal approach where $g_c = 1$. The charmed hadrons are, in the SHMc, all formed at the phase boundary, i.e. at hadronization, in the same way as all (u,d,s) hadrons. 

In Fig.~\ref{fig:HRG} it can be seen that, with that choice, the measured yield for J/$\psi$ mesons is very well reproduced, the uncertainty in the prediction is mainly caused by the uncertainty in the total charm cross section in Pb--Pb collisions. The enhancement visible for the J/$\psi$ meson compared to the purely thermal prediction equal a factor $g_c^2$ then , since the J/$\psi$ contains two charm quarks.  We note here that, because of the formation from deconfined charm quarks at the phase boundary, charmonia are unbound inside the QGP but their final yield exhibits enhancement compared to expectations using collision scaling from pp collisions, contrary to the original predictions based on ~\cite{Matsui:1986dk}. For a detailed discussion see ~\cite{Andronic:2017pug}.

For the description of yields of charmonia, feeding from excited charmonia is very small because of their strong Boltzmann suppression. For open charm mesons and baryons, this is not the case and feeding from excited $D^*$ and $\Lambda_c^*$ hadrons is an essential ingredient for the description of open charm hadrons ~\cite{Andronic:2021erx,Braun-Munzinger:2024ybd}. Even though the experimental delineation of the mass spectrum of excited open charm mesons and baryons is currently far from complete, the prediction of yields for D-mesons and $\Lambda_c$ baryons compares very well with the measurements\footnote{For $\Lambda_c$ baryons on has to augment the currently measured charm baryon spectrum with additional states to achieve complete agreement with experimental data~\cite{Andronic:2021erx,Braun-Munzinger:2024ybd}.}, both concerning transverse momentum and centrality dependence. 

The successful prediction  of yields of hadrons with more than 1 charm quark is also direct evidence that charm quarks are deconfined inside the fireball formed in nuclear collisions at LHC energy. The typical size of this deconfined system has linear dimension of order 10 fm. The apparent absence of string-like forces with typical range of about 1 fm inside this hot system is fully consistent with the presence of deconfined charm quarks and indicates that also all light quarks could be deconfined. The topic of deconfinement in high energy nuclear collisions is also discussed in ~\cite{Gross:2022hyw,Andronic:2017pug}.

\begin{figure}[!htb]
    \centering
    \includegraphics[width=0.65\linewidth,clip=true]{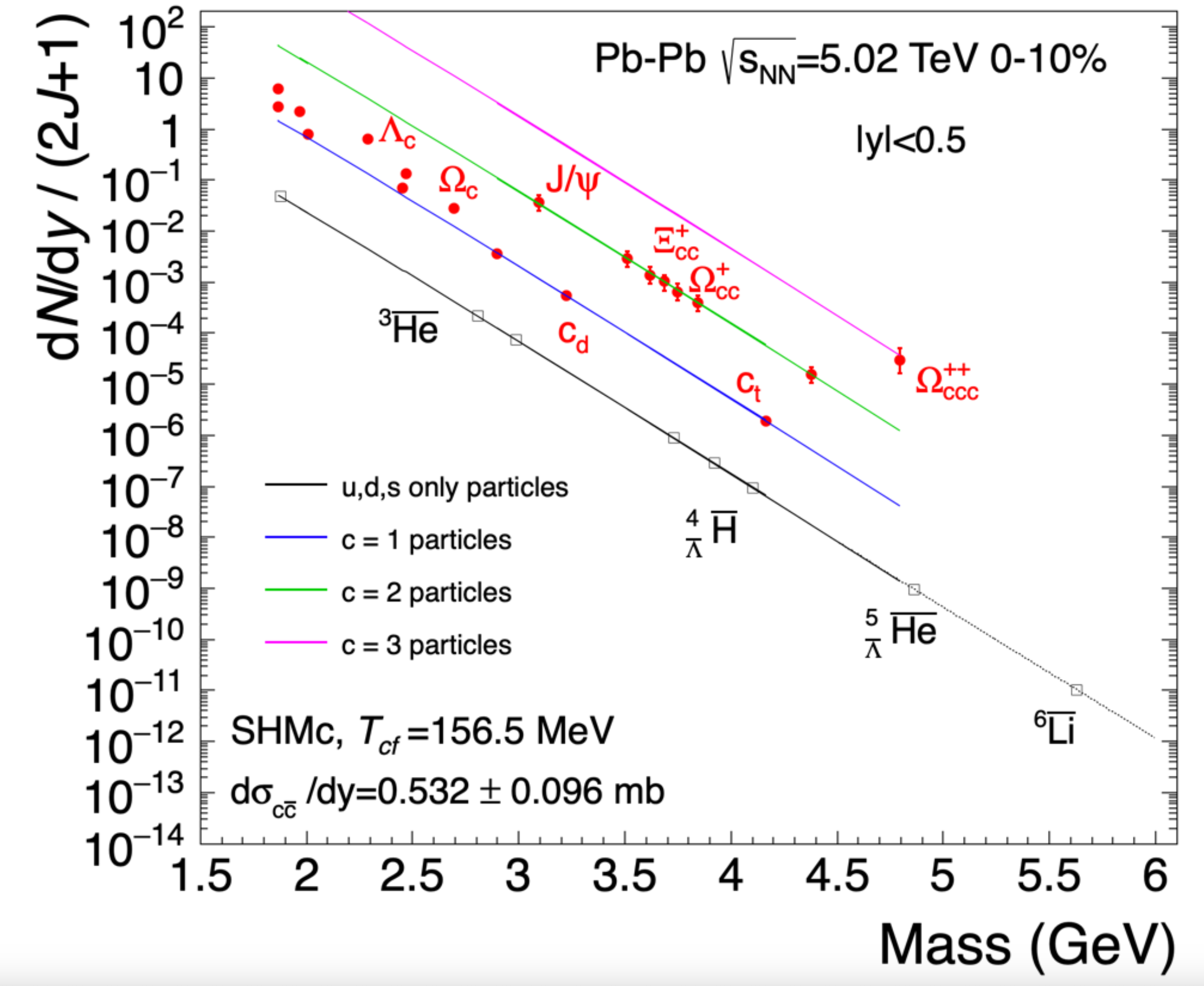}
    \caption{Predictions of yields for singly- and multiply-charmed hadrons by the Statistical Hadronization Model with charm SHMc. Figure from~\cite{Andronic:2021erx} and adapted to include also selected light nuclei and hyper-nuclei.}
    \label{fig:hierarchy}
\end{figure}

A quite spectacular hierarchy of enhancements then emerges if one compares the predicted yields of single, double and triple charm hadrons. This is exhibited in Fig.~\ref{fig:hierarchy}.

Future measurement campaigns at the LHC  in Run 3 and Run 4 and with ALICE 3 ~\cite{ALICE:2022wwr} will yield detailed information on the production cross sections of hadrons with multiple charm quarks as well as excited charmonia.  We expect that expe\-ri\-men\-tal tests of these predictions will lead to a more fundamental understanding of confinement/deconfinement and hadronization. Note that hadrons with two or more charm quarks cannot be produced in a single nucleon-nucleon collision. Confirmation of the SHMc prediction for multi-charm hadrons would imply that also the hadronization of these exotic hadrons can be described with a single parameter, i.e. the (pseudo-)critical temperature of the QCD cross-over phase transition.

An instructive way to proceed  is obtained by analyzing the centrality dependence of the yield ratio $D^0/(J/\psi)$ for Pb--Pb collisions and comparing the results to the predictions of the SHMc. These $D^0$ and $J/\psi$ production cross sections have for the 1st time been measured down to $p_t$ = 0. The experimental yield ratio $D^0/(J/\psi)$ is  in very good agreement with the SHMc prediction, as demonstrated in Fig.~\ref{fig:DotoJpsi}. This result buttresses the interpretation that open and hidden charm states are both produced by statistical hadronization at the phase boundary. An informative comparison between SHMc predictions  and data for a number of open charm hadrons is shown in~\cite{Andronic:2021erx}.

For center-of-mass energies $\sqrt{s_{NN}}\geq$ 10 GeV the chemical  freeze-out parameters closely agree with the phase boundary line obtained from lQCD ~\cite{HotQCD:2018pds,Borsanyi:2020fev}. We conclude that hadronization from a equilibrated fireball is independent of particle species and only dependent on the values of $T$ and $\mu_B$ at the phase boundary. At LHC energies where the chemical potential vanishes with good precision ~\cite{ALICE:2023ulv},  only  the freeze-out temperature $T = T_{pc}$ is needed to describe hadronization.
      
\section{Short remarks on charm quarks and the QGP equation of state}

The role of charm  degrees of freedom on the hot matter equation of state has recently been investigated using the tools of lQCD ~\cite{Bazavov:2014yba,Bazavov:2023xzm}. Since charm quarks are heavy, with a mass near 1.27 GeV (at a scale of 2 GeV ~\cite{ParticleDataGroup:2024cfk}) they are very rare. In a hot fireball at temperature $T = T_{pc}$, and in full grand-canonical equilibrium, as is assumed in lQCD calculations, the number of charm  quarks in a fireball with volume of 4000 $\mathrm{fm}^3$ per unit rapidity is 0.03, rising to 3.3 at $T = 2 T_{pc}$. At that level, the charm quarks make negligible contributions to the energy density, and hence, equation of state of the QGP. However, in high energy nuclear collisions, the large majority of charm quarks in the fireball is not created thermally, but in initial hard collisions ~\cite{Braun-Munzinger:2000csl,Andronic:2017pug} and subsequently thermalized in the hot QGP. For a central Pb-Pb collisions at $\sqrt{s_{NN}} = {5.02}$ TeV this translates into about 13 charm quark pairs per unit rapidity ~\cite{Andronic:2021erx}. This implies that about 3.5 \% of the critical energy density is contained in charm quarks, not large, but maybe not negligible anymore. Whether the `quenched dynamical quarks' approximation underlying the work of ~\cite{Bazavov:2014yba} is applicable here is also an open question. For conditions expected for the Future Circular Collider FCC there will be further increased initial charm production and likely substantial thermal charm production ~\cite{Zhou:2016wbo,Dainese:2016gch} which could lead to large (of order 20 - 30 \%) contributions by charm quarks to the QGP energy density and imply significant changes in the QGP equation of state. It would be very interesting to add the appropriate `non-thermal' number of charm quarks into lQCD calculations to test this.

\section{Summary}

The pioneering papers ~\cite{Pomeranchuk:1951ey,Fermi:1953zz,Landau:1953gs} of the 1950s provided the basis for the first phenomenological investigations of early data on hadron production in high energy pp collisions~\cite{Hagedorn:1965st} based on thermodynamical concepts. Since these very early days many experimental and theoretical/phenomenological analyses and publications have demonstrated the importance and usefulness of the statistical hadronization models which emerged over the course of the past 50 years. For a rather concise summary of the literature see ~\cite{Andronic:2017pug}. We note here in particular two examples of the use of SHM and SHMc to provide insight into the physics of the QGP. In the year 2006 two papers appeared nearly simultaneously (but based on separate investigations) on the energy dependence of thermodynamic parameters ~\cite{Andronic:2005yp,Cleymans:2005xv}. These papers together have to-date (May 2025) more than 1700 citations and provide the basis for many quantitative  QGP studies. In the year 2021 a paper appeared on multi-charm hadron production and the mechanism of hadronization. This paper has laid the basis for part of the physics program of the new ALICE 3 ~\cite{ALICE:2022wwr} experiment at the `High Luminosity  LHC'. 

Of course there are also physics results which cannot be directly obtained from the SHM or SHMc. To compute, in addition to particle multiplicities, also the corresponding transverse momentum distributions or particle flow coefficients one needs to connect SHM with dynamical models. This has recently been achieved ~\cite{Andronic:2023tui} by coupling the SHM(c) to models based on relativistic hydrodynamics, with good success. However, the thermal approach underlying SHM clearly breaks down for systems far off thermal equilibrium. Particle production at high transverse momentum, such as jet production, or processes involving energy loss of quarks and gluons in the dense fireball formed in high energy nuclear collisions cannot be described in the thermal framework outlined here.  Surprisingly, the thermal approach works well to understand the particle multiplicities within a given jet, even for jets originating from e$^+$e$^-$ collisions ~\cite{Becattini:1995if,Andronic:2008ev,Becattini:2008tx}.

\section{Acknowledgments} 
We acknowledge continued and long-term collaboration with Anton Andronic on many topics described in this contribution. This work is supported by the DFG Collaborative Research Centre SFB 1225 (ISOQUANT). PBM thanks Anar Rustamov for important discussions and help with the tex file. K.R. acknowledges the Polish National Science Centre (NCN) support under OPUS Grant No. 2022/ 45/B/ST2/01527 and of the Polish Ministry of Science and Higher Education. K.R. also acknowledges fruitful discussions with B. Friman, F. Karsch, P.M. Lo and C. Sasaki.

\bibliographystyle{unsrt}

\bibliography{PBM_KR_JS}
\end{document}